\documentclass[reprint,showpacs,superscriptaddress,amsmath,amssymb,aps]{revtex4-1}

\usepackage{graphicx}
\usepackage{color}
\usepackage{dcolumn}
\usepackage{bm}
\usepackage{amsmath}
\usepackage{amsfonts}
\usepackage{amssymb}
\usepackage{float}
\usepackage[font=small]{caption}
\usepackage[colorlinks,linkcolor=blue, citecolor=blue, urlcolor=blue]{hyperref} % to click on a reference number and go directly to it.

\begin{document}
\restylefloat{figure} % this, I think, will reduce spaces between images and text.
\title{Magnetoelectric properties of the multiferroic CuCrO$\bf_2$ studied by means of \textit{ab initio} calculations and Monte Carlo simulations}

\author{Ahmed Albaalbaky}%A. Albaalbaky
\altaffiliation{ahmed.baalbaky@hotmail.com} 
\affiliation{Normandie Universit\'e, UNIROUEN, INSA Rouen, CNRS, GPM, 76800 Saint \'Etienne du Rouvray, France}
\author{Yaroslav Kvashnin} %Y.O. Kvashnin
\affiliation{Department of Physics and Astronomy, Division of Materials Theory, Uppsala University, Box 516, SE-75120 Uppsala, Sweden}
\author{Denis Ledue}%D. Ledue
\affiliation{Normandie Universit\'e, UNIROUEN, INSA Rouen, CNRS, GPM, 76800 Saint \'Etienne du Rouvray, France}
\author{Renaud Patte}%R. Patte
\affiliation{Normandie Universit\'e, UNIROUEN, INSA Rouen, CNRS, GPM, 76800 Saint \'Etienne du Rouvray, France}
\author{Raymond Fr\'esard}%R. Fr\'esard
\affiliation{Normandie Universit\'e, UNICAEN, ENSICAEN, CNRS, CRISMAT, 14050 Caen, France}

\begin{abstract}
Motivated by the discovery of multiferroicity in the geometrically frustrated triangular antiferromagnet CuCrO$_2$ below its N\'eel temperature $T_N$, we investigate its magnetic and ferroelectric properties using \textit{ab initio} calculations and Monte Carlo simulations. Exchange interactions up to the third nearest neighbors in the $ab$ plane, inter-layer interaction and single ion anisotropy constants in CuCrO$_2$ are estimated by series of density functional theory calculations. In particular, our results evidence a hard axis along the [110] direction due to the lattice distortion that takes place along this direction below $T_N$. Our Monte Carlo simulations indicate that the system possesses a N\'eel temperature $T_N\approx27$ K very close to the ones reported experimentally ($T_N = 24-26$ K). Also we show that the ground state is a proper-screw magnetic configuration with an incommensurate propagation vector pointing along the [110] direction. Moreover, our work reports the emergence of spin helicity below $T_N$ which leads to ferroelectricity in the extended inverse Dzyaloshinskii-Moriya model. We confirm the electric control of spin helicity by simulating $P$-$E$ hysteresis loops at various temperatures. 
\end{abstract}

\date{\today}

\maketitle

\section{\label{sec:introduction}Introduction}

\begin{figure}[b]
\centering
\includegraphics[width=6cm]{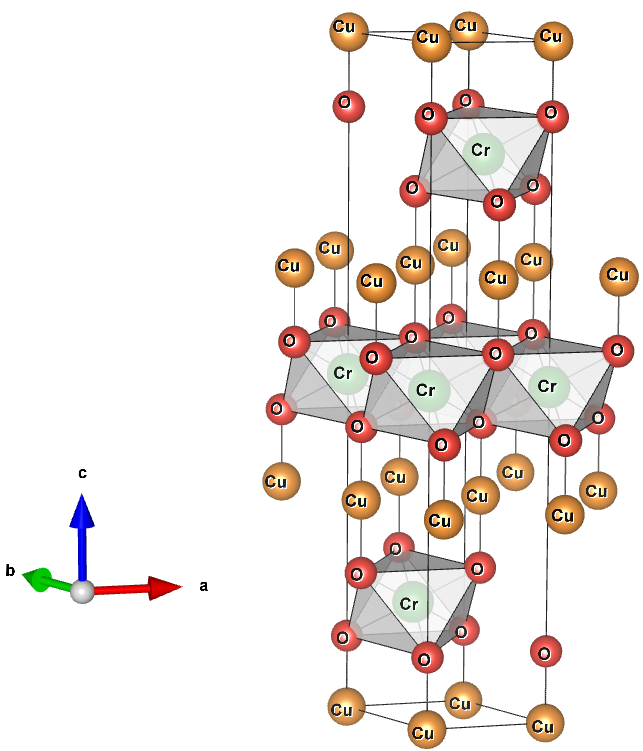}
\caption{(Color online) Delafossite structure of CuCrO$_2$.}
\label{Fig:struc}
\end{figure}

Through the discovery of the mineral CuFeO$_2$ in 1873, Friedel opened the door to the delafossites ABO$_2$ \cite{Fri73,Sha71}. Such a family crystallizes in the layered $R\bar{3}m$ space group, see
Fig.~\ref{Fig:struc}. The diversity of properties they exhibit raises up an ever increasing interest in this class of compounds. In particular, the discovery of simultaneous transparency and $p$-type conductivity in CuAlO$_2$ by Kawazoe \textit{et al.} \cite{Kaw97}, laid ground for the development of transparent optoelectronic devices. Furthermore, depending on the chemical composition, a plethora of behaviors can be evidenced. For instance, for A in a $d^9$ configuration, e.g., A = Pd or Pt, highly metallic compounds with anomalous temperature dependence of the resistivity have been reported \cite{Tak07,Hic12,Hic15,Kus15}. The transport in these compounds has been found to be strongly anisotropic, with a degree of anisotropy that may reach 1000 \cite{Tak07,Hic12,Daou15}. For A in a $d^{10}$ configuration, the semi-conducting materials CuBO$_2$, with B = Cr, Fe, Rh, may be turned into promising thermoelectric ones through hole doping \cite{Oku05,Noz07,Kur06} --- in particular, an especially high power factor has been found in the case of CuRh$_{1-x}$Mg$_x$O$_2$ \cite{Mai09b}, which transport coefficients served as a basis for the Apparent Fermi Liquid
scenario \cite{Kre12}. Regarding the magnetic compounds CuFeO$_2$ and CuCrO$_2$, many studies point towards a strong coupling of the magnetic and structural degrees of freedom \cite{Mek92,Mek93,Pet00,Kimura2006,Ye06,Eye08b,Soda2009,Mai09a,Jia12}, that paves the way to multiferroelectricity.

\begin{figure}[tb]
	\centering
	\includegraphics[width=8cm]{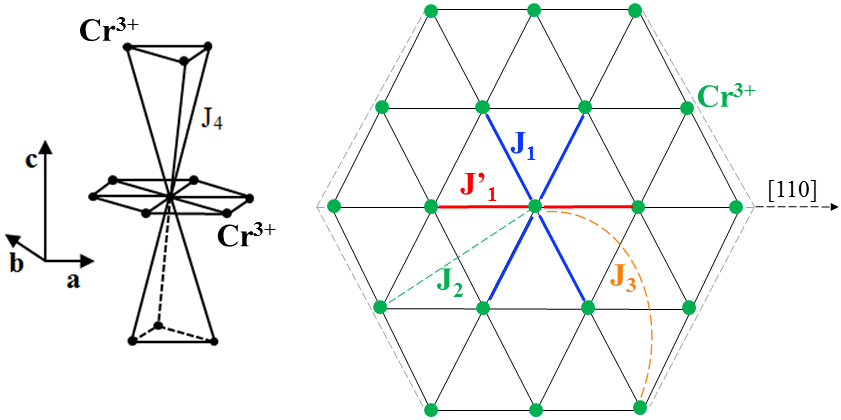}
	\caption{(Color online) Interlayer and intralayer exchange interactions within an $ab$ plane (blue bonds correspond to $J_1$ and red bonds correspond to $J'_1$ with $J_1/J'_1<1$).}
	\label{Fig:disJ}
\end{figure}

With its frustrated triangular lattice CuCrO$_2$ received a lot of attention since it is ferroelectric without applying magnetic fields or doping upon Cr$^{3+}$ sites, unlike CuFeO$_2$ \cite{Kimura2006,Haraldsen2010}. The emergence of ferroelectricity in CuCrO$_2$ is induced by the proper-screw magnetic ordering below the N\'{e}el temperature $T_N$, and the control of this ferroelectricity by an applied magnetic field is very important for new spin-based device applications. CuCrO$_2$ forms a rhombohedral lattice where the edge-shared CrO$_6$ layers are alternatively stacked between Cu$^+$ layers along the c-axis as shown in Fig.~\ref{Fig:struc}. Due to the weak inter-layer interaction $J_4$ (Fig.~\ref{Fig:disJ}), the material behaves as a quasi-2D magnet, which makes it even more interesting.
The magnetic properties of CuCrO$_2$ have been investigated by neutron diffraction experiments \cite{Kadowaki,Soda2009,Soda2010,Frontzek2012,Poienar2009}. It was shown that the magnetic configuration of CuCrO$_2$ below $T_N$ is proper screw with an incommensurate propagation vector \textit{\textbf{q}} = (0.329, 0.329, 0) \cite{Poienar2009} pointing along the [110] direction.
Such deviation from the commensurate magnetic configuration of \textit{\textbf{q}} = (1/3, 1/3, 0) is due to the lattice distortion that takes place along the [110] direction below $T_N$ upon the spiral-spin ordering  which leads to anisotropic in-plane exchange interactions $J_1$ and $J'_1$ (Fig.~\ref{Fig:disJ}) \cite{Kimura2009}. Polarized neutron-diffraction measurements on single crystals of CuCrO$_2$ \cite{Soda2009} showed that the spins are oriented in a spiral plane parallel to the (110) plane suggesting that the [110] direction is a hard axis. 

The electric polarization emerges upon the spiral-spin ordering \cite{Seki,Soda2009,Kimura2008}, which reflects the strong coupling between non-collinear magnetic ordering and ferroelectricity in CuCrO$_2$. 
Within the spin-current model or the inverse Dzyaloshinskii-Moriya (DM) mechanism \cite{Tokura,Tokura_Review,Noriki2014}, the electric polarization $\mathbf{P}_{ij}$ produced between the canted spins $\mathbf{S}_i$ and $\mathbf{S}_j$, located at sites $i$ and $j$, respectively, is given by
\begin{equation}
\mathbf{P}_{ij} \propto  \mathbf{e}_{ij}\times(\mathbf{S}_i\times\mathbf{S}_j)\equiv\mathbf{p}_1
\label{DM model}
\end{equation}
where ${\bf e}_{ij}$ is a unit vector joining the sites $i$ and $j$. 
However, Eq.(\ref{DM model}) fails to explain the emergence of ferroelectricity in CuCrO$_2$ because in the proper-screw configurations, $(\mathbf{S}_i\times\mathbf{S}_j)$ is parallel to $\mathbf{e}_{ij}$ ($\mathbf{e}_{ij}$ is along the [110] direction due to symmetry considerations \cite{Seki}) unlike the cycloid spin structures. 

Based on symmetry considerations, Kaplan and Mahanti \cite{Kaplan_Mahanti} introduced an additional contribution $\mathbf{p}_2 \propto(\mathbf{S}_i\times\mathbf{S}_j)$ to the macroscopic polarization which contributes in both cycloid and proper-screw configurations. 
Therefore, within this model, now referred to as extended DM model, the total polarization is given by
\begin{equation}
\mathbf{P}=\mathbf{p}_1+\mathbf{p}_2
\label{Extended DM model}
\end{equation}

In this study, we investigate the magnetoelectric properties of CuCrO$_2$ by means of a combination of Density Functional Theory (DFT) calculations and Monte Carlo (MC) simulations. More precisely, we estimate a set of exchange interactions and anisotropy constants and confront it to the experimental magnetic properties and we verify the appearance of spiral spin ordering at low temperatures  which can be related to the ferroelectric polarization. 

In Sec.~\ref{sec:technique} we detail briefly the DFT method that we used to extract the coupling and anisotropy constants in CuCrO$_2$, while the model and MC method are presented in Sec.~\ref{sec:MC}. Sec.~\ref{sec:Res} is devoted to the results where we discuss the magnetic and ferroelectric properties of CuCrO$_2$. A conclusion is given in Sec.~\ref{sec:Concl}.
 
\section{\label{sec:technique}DFT computational method} 

We performed a series of DFT calculations using full-potential linear muffin-tin orbital (FP-LMTO) method as implemented in RSPt \cite{rspt-book} code. An experimental crystal structure \cite{crstruc} was considered, taking into account a small in-plane lattic distortion, suggested in Ref.~\cite{Kimura2009}. Our results are in-line with earlier calculations \cite{Mai09a}. The DFT+$U$ \cite{lsdau} approach was used in order to take into account the effect of strong correlations between Cr $3d$ electrons. The adopted values of Hubbard $U$ and Hund’s exchange $J$ were 2.3 and 0.96 eV, which were extracted from first-principles calculations for a similar system LiCrO$_2$ \cite{Mazin2007}. The same computational scheme was used in a prior study on the magnetic properties of CuCrO$_2$ \cite{Jia12}. The Fully Localized Limit (FLL) \cite{FLL-DC} formed of the double-counting correction was applied. We calculated the exchange parameters between Cr$^{3+}$ ions by means of the magnetic force theorem \cite{lichtenstein,exch2}. The so-called muffin-tin head projection scheme was applied to construct the set of localized Cr-$d$ orbitals (for more details see Ref.~\cite{rspt-jijs}). The $J_{ij}$'s were extracted from both ferromagnetic and antiferromagnetic configurations. The obtained values turned out to be insensitive to the assumed magnetic order, which implies that they can be used as fixed parameters in a Hamiltonian describing the interacting spins. The spin-orbit coupling was taken into account only for the calculation of the magnetocrystalline anisotropy, which was calculated directly from the total energies.

\section{\label{sec:MC}Model and Monte Carlo simulation}

To model the magnetic properties of CuCrO$_2$, we note that Cr$^{3+}$ ions with $S$ = 3/2 spins are large enough to be treated classically, so we used the following classical three dimensional (3D) Heisenberg Hamiltonian 
\begin{align}
H  =& -\sum_{\langle i,j\rangle}J_{ij}\mathbf{S}_i\cdot\mathbf{S}_j - D_x\sum_iS_{ix}^2 -
D_z\sum_iS_{iz}^2 \nonumber \\
& + g\mu_B\mathbf{B}\cdot\sum_i\mathbf{S}_i 
\label{Hamiltonian}
\end{align} 
where $J_{ij}$ refers to the exchange interactions up to the 4$^{th}$ neighbors (Fig.~\ref{Fig:disJ}). The $x$-axis corresponds to the [110] direction and the $z$-axis corresponds to the [001] direction. $D_x<0$ and $D_z>0$ correspond to the hard and easy axes anisotropy constants respectively. The fourth term corresponds to the Zeeman energy where $\mathbf{B}$ is the applied magnetic field ($\mu_B$ is the Bohr magneton and $g=2$ is the Land\'{e} factor). 

To model the ferroelectric properties of CuCrO$_2$ and the coupling between the spins and the electric field $\mathbf{E}$, we added the following term to the previous Hamiltonian 
\begin{equation}
H_e= - A_0\mathbf{E}\cdot \sum_{{\langle i,j\rangle}}\mathbf{S}_i\times\mathbf{S}_j
\label{H:ferro}
\end{equation}
where the sum runs over the magnetic bonds along the [110] direction, and $A_0$ is a coupling constant related to the spin-orbit and spin exchange interactions. Adding this contribution leads to the model for multiferroics proposed by Kaplan and Mahanti \cite{Kaplan_Mahanti}.  

Our MC simulations \cite{MC} were performed on 3D triangular lattices (Fig.~\ref{Fig:struc} with only Cr$^{3+}$ ions) with periodic boundary conditions (PBC) using the standard Metropolis algorithm \cite{Metropolis} and the time-step-quantified method \cite{T.S.MC} when needed. 

Typically, the first $2\times10^4$ MC steps were discarded for thermal equilibration before averaging over the next $3\times10^5$ MC steps. Note that our results are averaged over 24 simulations with different random number sequences so that statistical fluctuations are negligible.

\section{\label{sec:Res}Results and discussions}

It was reported in Ref.~\cite{Kimura2009} that the lattice undergoes a tiny in-plane distortion $d=(a_2-a_1)/a_1$ below $T_N$ with $a_1$ and $a_2$ being the lattice constants along the [110] and the [100] directions, respectively. As a first step, we considered $d=0.0001$ \cite{Kimura2009} to calculate the exchange interactions and anisotropy constants in CuCrO$_2$. The extracted values given in Table~\ref{table1} (line 1) are very close to the ones reported in Ref.~\cite{Yamaguchi2010} concerning $J_1$ and $J'_1$ as well as the single ion anisotropy constants. Note that here $J_1/J'_1$ is very close to 1 ($J_1/J'_1=0.995$). Knowing that PBC favors the commensurate configuration when $J_1/J'_1$ is close to 1, large enough sizes are required to obtain an incommensurate magnetic ground state (GS).   
\begin{table}[b]
	\setlength{\extrarowheight}{3pt}
	\caption{Estimated DFT values of the exchange interactions and anisotropy constants (in meV). More precisely, for $d$=0.0001, the calculated value of $D_x$ was smaller than 10$^{-4}$ meV, which is negligible.}
	\begin{tabular}{*{8}{c}}  
		\hline
		\hline
		$d$ & $J'_1$ & $J_1$ & $J_2$ & $J_3$ & $J_4$ & $D_x$ & $D_z$\\
		\hline
		0.0001 & -2.419 & -2.407 & 0.012 & -0.266 & -0.060 & 0.000 & 0.033\\
		%\hline
		0.003 & -2.709 & -2.383 & 0.012 & -0.266 & -0.060 & -0.001 & 0.033\\
		\hline
		\hline
	\end{tabular}
	\label{table1}	
\end{table}
However, a MC simulation with 90$\times$90$\times$2 unit cells was not able to reproduce an incommensurate GS with this set of interactions ($d=0.0001$). Thus larger sizes of the simulation box were required which are not accessible within reasonable computer time \cite{Ahmed}. Therefore we enhanced the lattice distortion by a factor of 30 (i.e. $d=0.003$). We found that the new set of $J_{ij}$'s ($J_1/J'_1=0.88$) and anisotropy constants (Table~\ref{table1}) is a good candidate to reproduce an incommensurate GS for a system of reasonable size 45$\times$45$\times$2 unit cells. It is worth noting that the considered distortion mainly affect the first nearest neighbors interactions while the remaining interactions are  not affected. Also it is very interesting to note that the magnitude of the in-plane anisotropy constant ($D_x$) increases when enhancing the lattice distortion reflecting that this anisotropy results from the lattice distortion. 

\subsection{Magnetic properties}

\begin{figure}[tb]
	\centering
	\includegraphics[width=6cm]{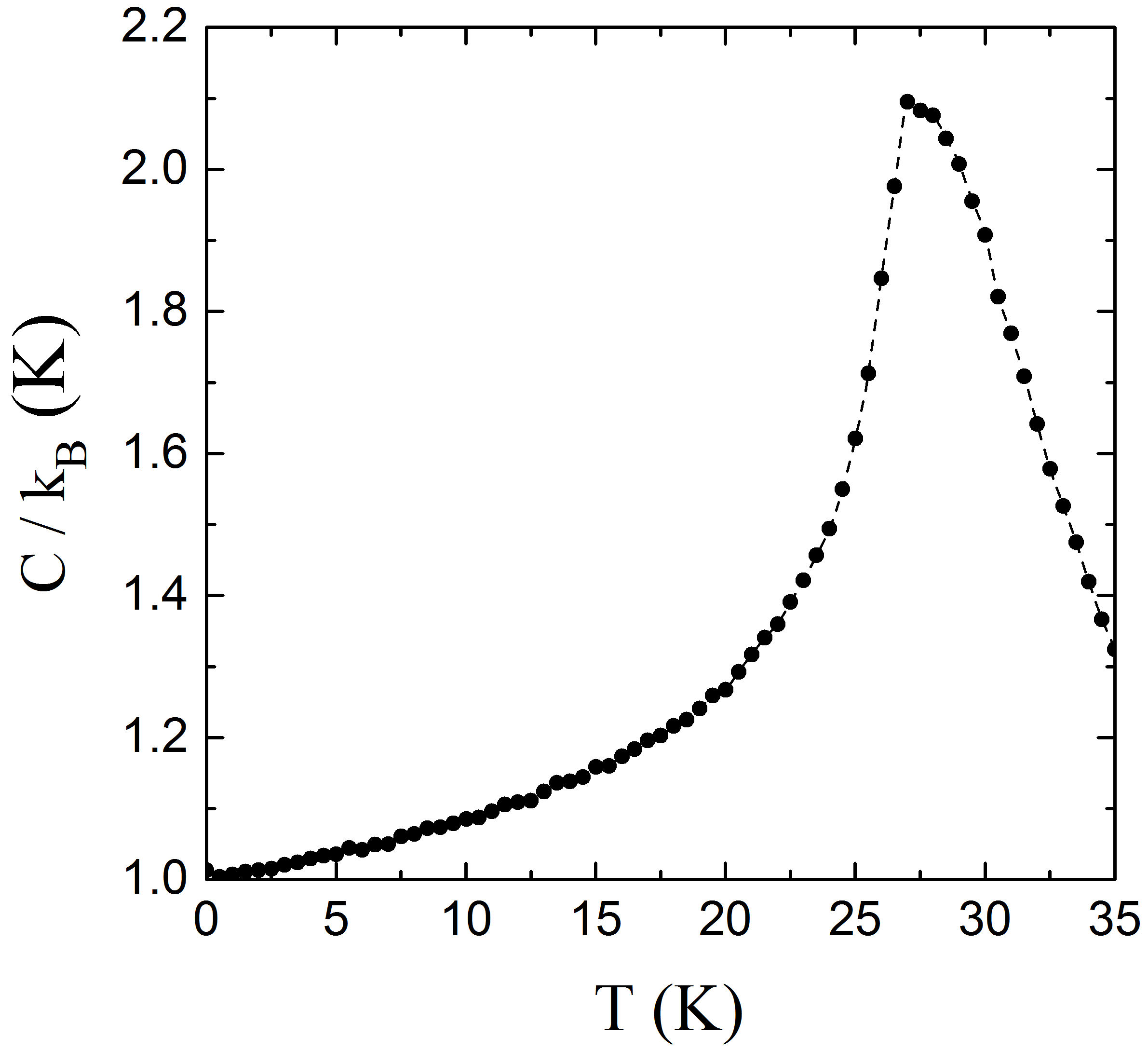}
	\caption{Simulated temperature dependence of the specific heat per spin of CuCrO$_2$. The parameter values are given in Table~\ref{table1} for $d$=0.003.}
	\label{Fig:Cv}
\end{figure}
In order to characterize the GS configuration and to estimate the N\'eel temperature $T_N$ we performed a first set of simulations without applying an external magnetic field. The following procedure has been retained: we started the simulations from random spin configurations at a high enough temperature ($T> T_N$) and we then cooled down to $T_{\rm final}$ = 0.01~K with a constant temperature step $\Delta T$ =
0.5~K.  

In order to estimate the N\'eel temperature, we calculated the specific heat per spin defined as 
\begin{equation}
C=\frac{1}{N}\frac{\partial U}{\partial T}=\frac{\langle E^2\rangle_T -\langle E\rangle^2_T}{Nk_BT^2} 
\label{Specific heat}
\end{equation}
where $U(T)=\langle E \rangle_T$ with $E$ being the energy of each magnetic configuration, $\langle\dots\rangle_T$ means thermal average, $N$ is the number of spins and $k_B$ is the Boltzmann constant. For the parameter set given in Table~\ref{table1} ($d=0.003$) the phase transition as signaled by the peak of the specific heat (Fig.~\ref{Fig:Cv}) takes place at $T_N=27.0\pm0.5$~K. This value is in a good agreement with the reported experimental values ($T_N = 24-26$ K) \cite{Oku05,Seki,Poienar2009}. This may be taken as a first validation of the extracted exchange interactions of Table~\ref{table1}.  
\begin{figure}[tb]
	\centering
	\includegraphics[width=6cm]{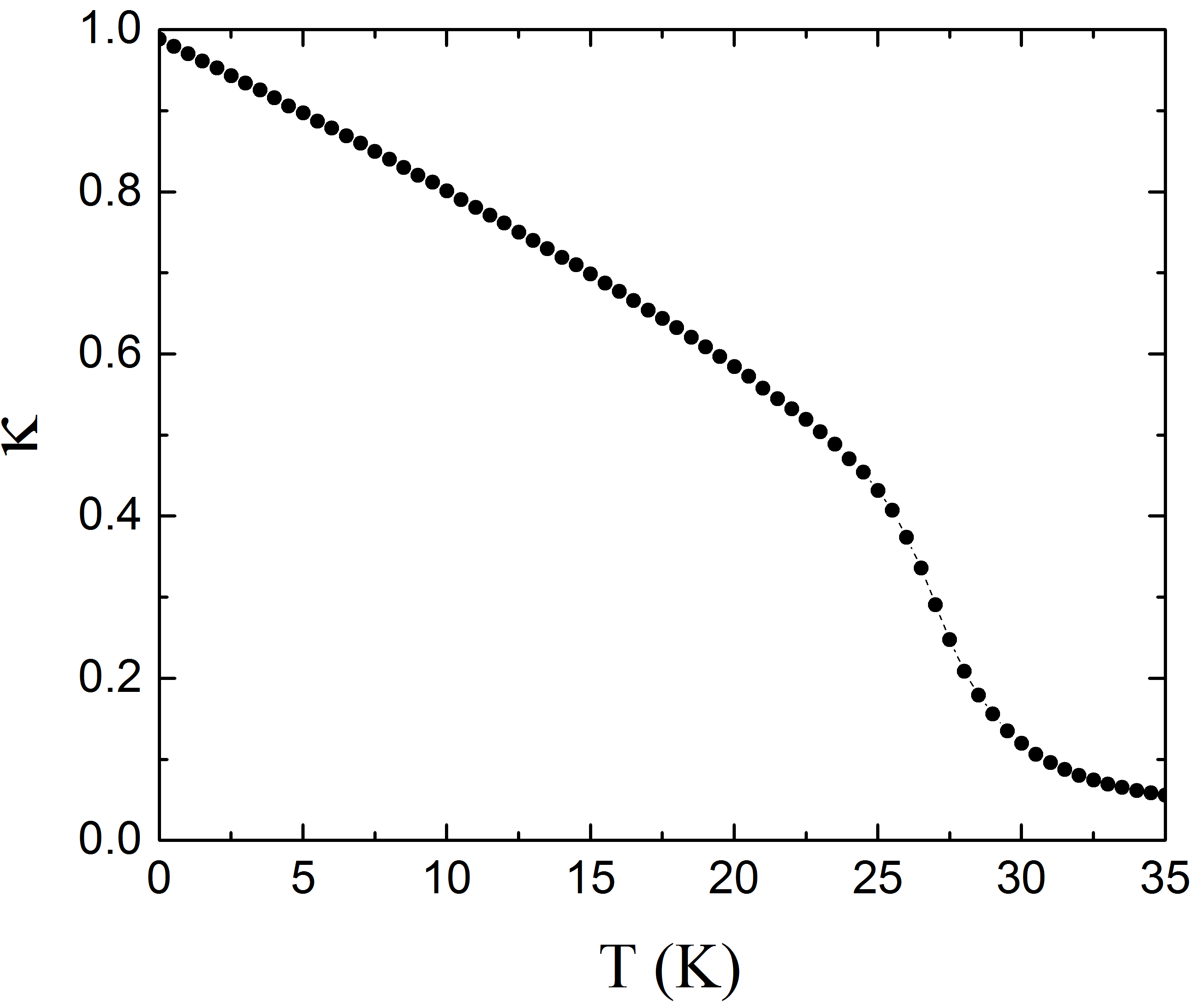}
	\caption{Simulated temperature dependence of the order parameter in CuCrO$_2$ (at $T\approx 0~$K, $\kappa \approx 0.988$).}
	\label{Fig:kappa}
\end{figure}

To characterize the nearly $120^\circ$ GS configuration we considered the
spin chirality defined as
\begin{equation}
\bm{\kappa}_p=\frac{1}{S^2}\frac{2}{3\sqrt{3}}(\mathbf{S}_1\times\mathbf{S}_2+\mathbf{S}_2\times\mathbf{S}_3+\mathbf{S}_3\times\mathbf{S}_1)
\label{kappa}
\end{equation}
\begin{figure}[b]
	\centering
	\includegraphics[width=7cm]{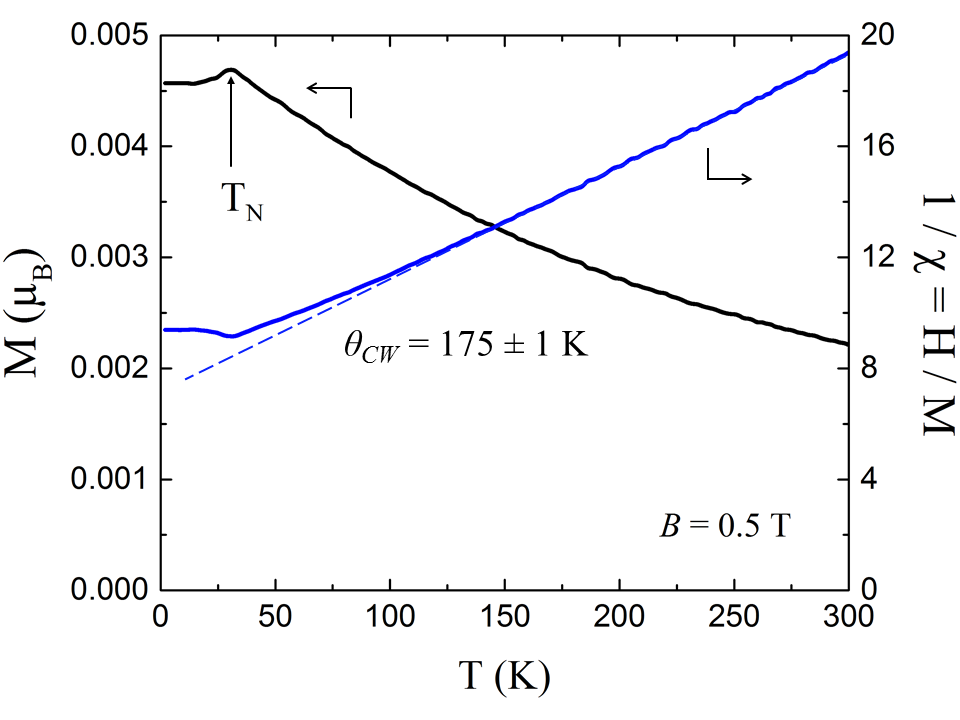}
	\caption{(Color online) Simulated temperature dependence of the magnetization per spin and the inverse susceptibility under $B$ = 0.5 T magnetic field in CuCrO$_2$.}
	\label{Fig:CW}
\end{figure}where 1, 2 and 3 refer to the spins at the corners of each elementary triangular plaquette $p$ in an $ab$ plane. Then we defined the order parameter per plane to be $\lambda = \frac{1}{n_p} \|\sum_{p} \bm{\kappa}_p\|$ where $n_p$ is the number of plaquettes per plane, and finally the order parameter of the whole system was defined as $\kappa=\langle \bar{\lambda} \rangle_T$ where $\bar{\lambda}$ is the average of $\lambda$ over the $ab$ planes. We found that the direction of the vector chirality ($\bm{\lambda}$) of each $ab$ plane is pointing along the [110] direction confirming the fact that the spins are oriented in the (110) plane as reported in Ref.~\cite{Soda2009}. Fig.~\ref{Fig:kappa} shows the variation of the order parameter as function of temperature. At $T\approx0$ K, $\kappa\approx0.988$ indicates a small deviation from the commensurate (120$^\circ$) configuration of $\kappa=1$. Moreover, the simulated value of \textit{\textbf{q}} $\approx$ (0.322, 0.322, 0) confirms that the GS is an incommensurate configuration very close to the reported experimental configuration of \textit{\textbf{q}} $=$ (0.329, 0.329, 0) \cite{Poienar2009}. This good agreement may be taken as another validation of the parameters of Table~\ref{table1}.  

On the other hand, the magnetic field dependence of the magnetization calculated along the easy axis ($z$-axis) shows a linear behavior ($-5~T<H_z<5~T$) confirming the antiferromagnetic nature of the GS (not shown here). 

\begin{figure}[tb]
\centering
\includegraphics[width=6cm]{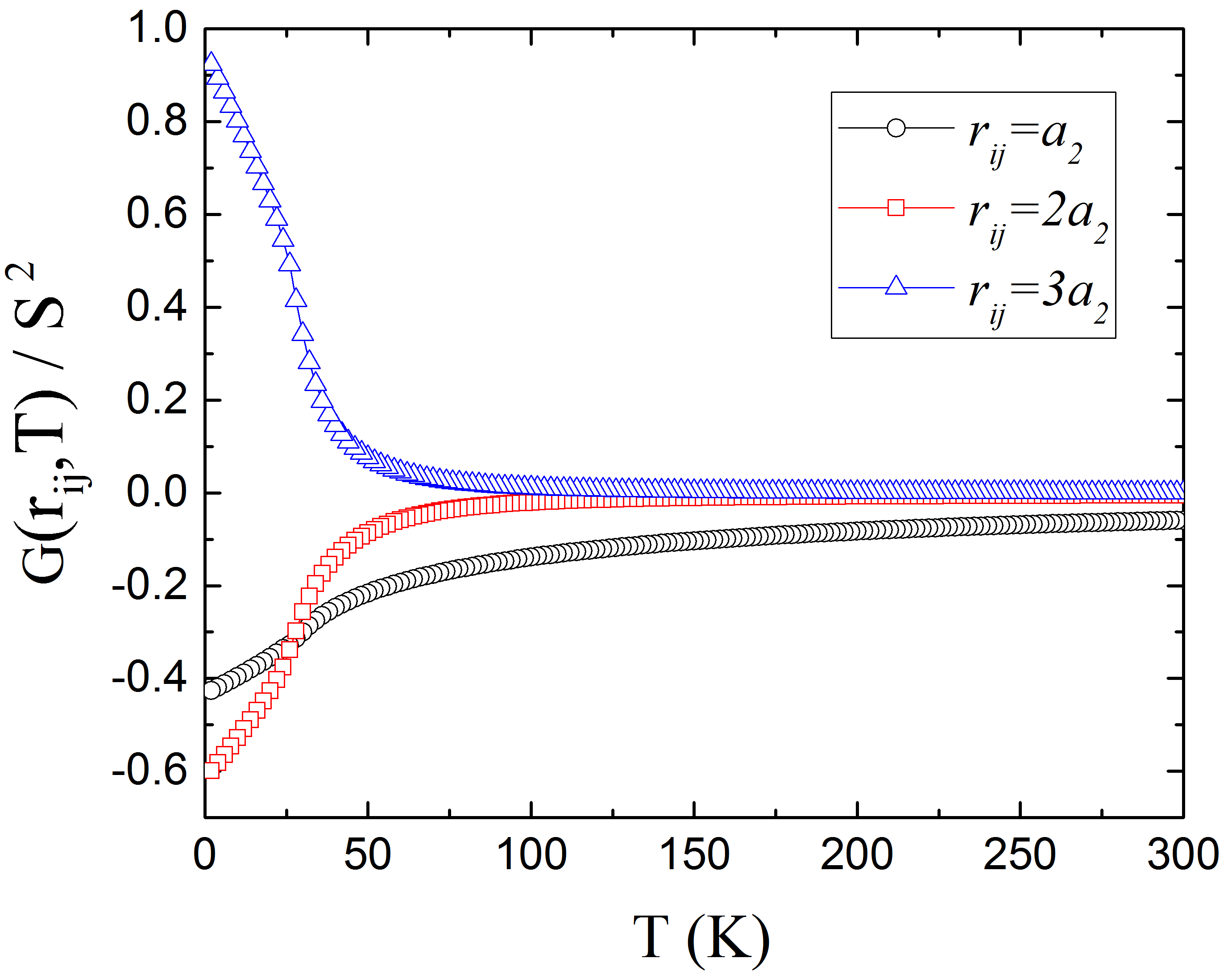}
\caption{(Color online) Simulated temperature dependence of the spin-spin correlation functions along [100] calculated at distances $a_2$ (black circles), $2a_2$ (red squares) and $3a_2$ (blue triangles) in CuCrO$_2$.} 
\label{Fig:Correlation}
\end{figure}

Magnetic properties under 0.5 T were simulated between 300 K and 2 K to estimate the Curie$-$Weiss temperature ($\theta_{CW}$). Fig.~\ref{Fig:CW} shows the variation of the magnetization and inverse susceptibility measured along the applied magnetic field. 
It can be seen that $1/\chi$ obeys well the Curie$-$Weiss law for antiferromagnets ($1/\chi = (T+\theta_{CW})/C$, with $C$ is the Curie constant) at high temperatures with $\theta_{CW} = 175~\pm~1$ K close to the measured experimental values ($\theta_{CW} = 160-170$ K) \cite{Oku05,Okuda2013}. The $1/\chi$ curve starts to deviate from the linear behavior at about 100 K. In order to understand the origin of this deviation we calculated the temperature dependence of the spin-spin correlation function defined as $G(r_{ij},T) = \langle\mathbf{S}_i\cdot\mathbf{S}_j\rangle_T$ along the [100] direction. As shown in Fig.~\ref{Fig:Correlation}, short-range antiferromagnetic correlations start to develop below $\sim$ 100~K, which leads to the deviation from the Curie$-$Weiss law seen in Fig.~\ref{Fig:CW}. Furthermore, these correlation functions exhibit inflection points close to $T_N$ estimated from the specific heat curve (Fig.~\ref{Fig:Cv}).
Besides, an anomaly in the magnetization curve (Fig.~\ref{Fig:CW}) appears at $28\pm2$ K consistent with the estimate of $T_N$ from the specific heat curve. We note that the ratio $\theta_{CW}/T_N \approx 6.5$ ($\gg 1$) reflects the frustrated nature of the GS \cite{frustration1,frustration2}.

\subsection{Ferroelectric properties}

\begin{figure}[tb]
	\centering
	\includegraphics[width=6.5cm]{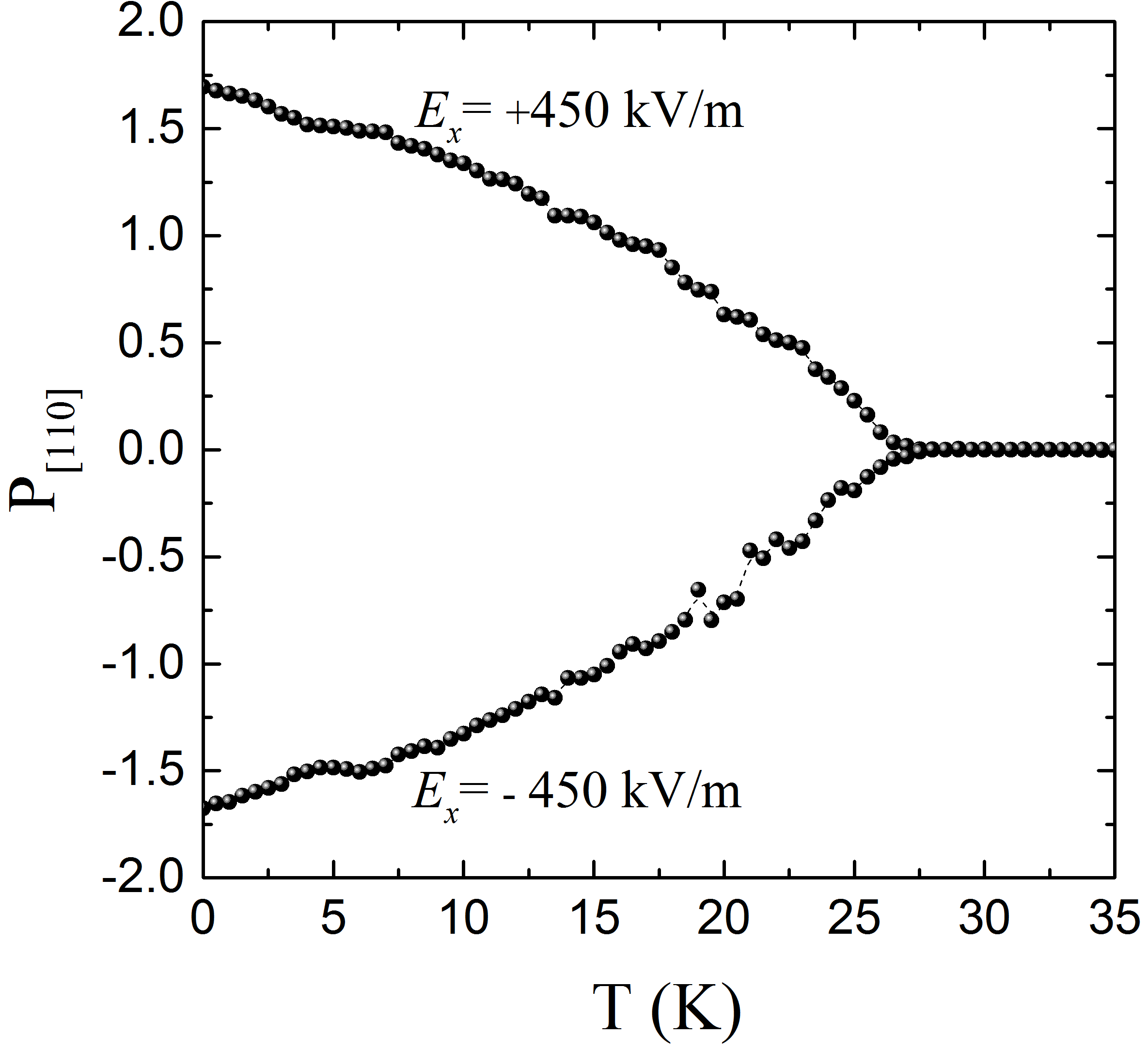}
	\caption{Simulated temperature dependence of the electric polarization \textbf{P} calculated along the [110] direction in CuCrO$_2$.}
	\label{Fig:Pol}
\end{figure}

In this section, we considered the Hamiltonian $H+H_e$. In these simulations, we applied a poling electric field during the cooling process to obtain a single ferroelectric domain. We then turned it off just before statistical averaging to calculate $\mathbf{p}_2$ which is associated to the spontaneous ferroelectric polarization (Eq.~(\ref{Extended DM model})) according to Ref.~\cite{Kaplan_Mahanti}. Fig.~\ref{Fig:Pol} shows the temperature dependence of $P_{[110]}$, the projection of $\mathbf{p}_2$ along the [110] direction, which starts to develop at $T_N$. It is clearly seen that by switching the poling electric field, $P_{[110]}$ can be reversed.

\begin{figure}[t]
	\centering
	\includegraphics[width=7cm]{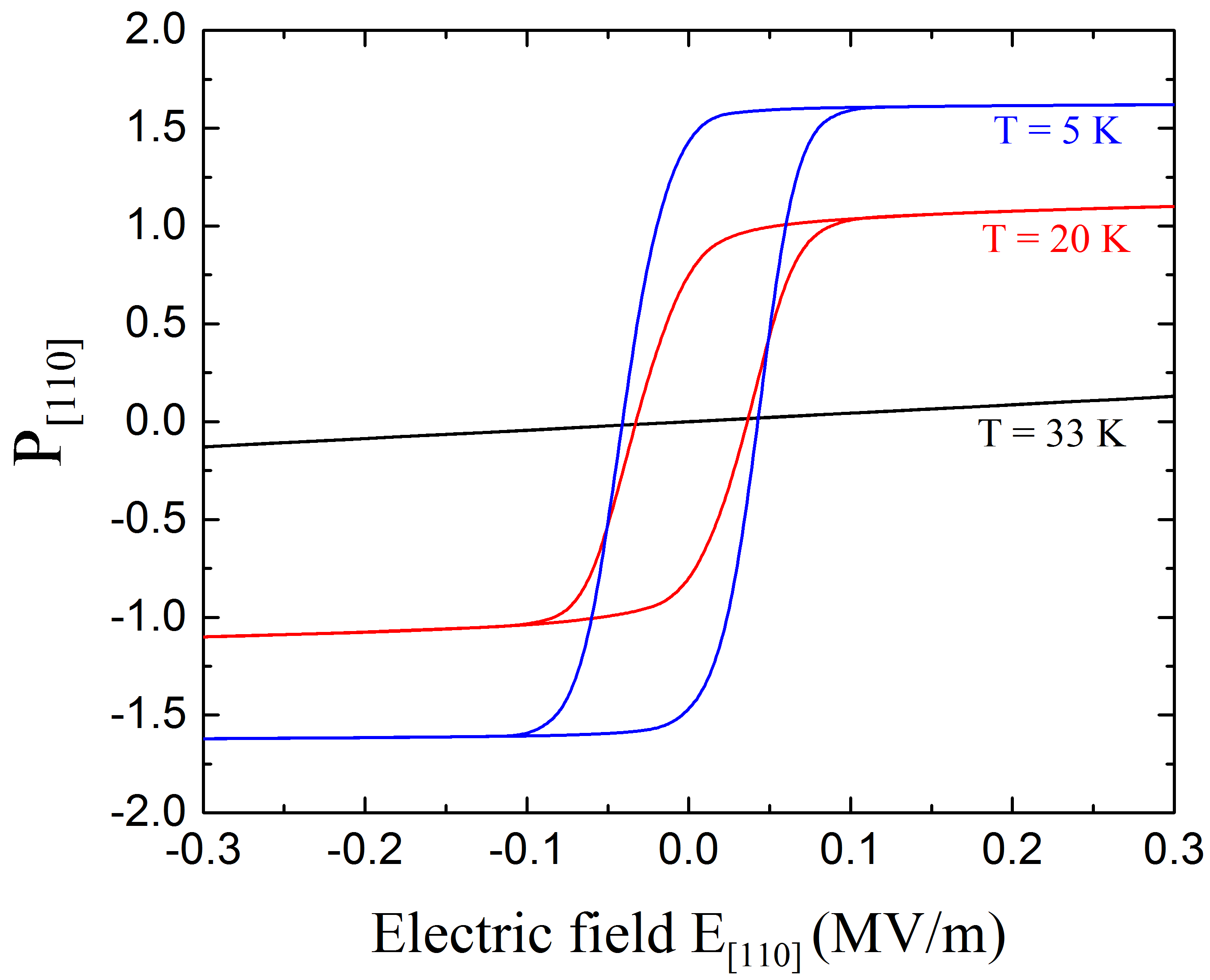}
	\caption{(Color online) $P$-$E$ hysteresis loops simulated at different temperatures in CuCrO$_2$.} 
	\label{Fig:hyst}
\end{figure}

Further insight into the degree of electrical polarization may be gained through the knowledge of the $P$-$E$ hysteresis loops, which are shown in Fig.~\ref{Fig:hyst} at different temperatures. $P_{[110]}$ shows a linear $E$ dependence without hysteresis above $T_N$ because the system is in the paraelectric phase, while clear hysteresis loops are seen for temperatures below $T_N$. This strongly suggests that ferroelectricity is induced by the out-of-plane incommensurate magnetic configuration, in agreement with Ref.~\cite{Kimura2008}.
Also, it can be seen that below $T_N$ the saturation field $E_{sat}\approx 8.9\times10^{-2}$ MV/m is independent of the temperature. The hysteresis loop simulated at $5$ K shows an electric coercive field for \textbf{$P_{[110]}$} reversal $E_r\approx4.2\times10^{-2}$ MV/m very close to that measured experimentally ($E_r=5.1\times10^{-2}$ MV/m \cite{Kimura2009_PRL}). Note that the reversal of $P_{[110]}$ results from the reversal of the helicity of each $ab$ atomic plane. Thus our simulations confirm the electric control of spin helicity in CuCrO$_2$ as reported in Ref.~\cite{Soda2009}.

\section{\label{sec:Concl}Conclusion}

In this paper, we proposed estimates of the exchange interactions and single ion anisotropy constants in the multiferroic CuCrO$_2$ using DFT calculations. They were checked against the experimental N\'eel and Curie$-$Weiss temperatures as well as the electric coercive field, thereby proving them to be good candidates to model the magnetoelectric properties of CuCrO$_2$. We showed that the lattice distortion that takes place below $T_N$ is responsible for the appearance of a weak in-plane hard-axis anisotropy. Regarding the magnetic properties, we obtained a peak in the specific heat curve at $T_N\approx27$ K very close to the experimental observations. Furthermore the ground-state has been shown to be an antiferromagnetic incommensurate proper-screw configuration. The estimated $\theta_{CW} \approx 175$ K is in a good agreement with experimental data too. Also, our simulated $P$-$E$ hysteresis loops confirm the electric control of spin helicity which is related to the ferroelectric polarization below $T_N$. 

\section*{Acknowledgments}

We gratefully thank M. Alouani and S. H\'ebert for stimulating discussions. We are grateful to the Centre R\'egional Informatique et d'Applications Numeriques de Normandie (CRIANN) where our simulations were performed as project number 2015004. We also acknowledge the computational resources provided by the Swedish National Infrastructure for Computing (SNIC) and Uppsala Multidisciplinary Center for Advanced Computational Science (UPPMAX). The authors acknowledge the financial support of the French Agence Nationale de la Recherche (ANR), through the program Investissements d'Avenir (ANR-10-LABX-09-01) and LabEx EMC3.


\begin{thebibliography}{0}
	
\bibitem{Fri73} C. Friedel, Sciences Academy \textbf{77}, 211 (1873).

\bibitem{Sha71} R. D. Shannon, D. B. Rogers and C. T. Prewitt, Inorg. Chem. \textbf{10}, 713 (1971); C. T. Prewitt, R. D. Shannon and D. B. Rogers, Inorg. Chem. \textbf{10}, 719 (1971); D. B. Rogers, R. D. Shannon, C. T. Prewitt and J. L. Gillson, Inorg. Chem. \textbf{10}, 723 (1971).

\bibitem{Kaw97} 
H. Kawazoe, M. Yasukawa, H. Hyodo, M. Kurita, H. Yanagi, and H. Hosono, Nature \textbf{389}, 939 (1997). 

\bibitem{Tak07} H. Takatsu, S. Y. Onezawa, S. M. Ouri, S. Nakatsuji, K. T. Anaka, and Y. Maeno, J. Phys. Soc. Jpn \textbf{76}, 104701 (2007).

\bibitem{Hic12} C. W. Hicks, A. S. Gibbs, A. P. Mackenzie, H. Takatsu, Y. Maeno, and E. A. Yelland, Phys. Rev. Lett. \textbf{109}, 116401 (2012).

\bibitem{Hic15} C. W. Hicks, A. S. Gibbs, L. Zhao, P. Kushwaha, H. Borrmann, A. P. Mackenzie, H. Takatsu, S. Yonezawa, Y. Maeno, and E. A. Yelland, Phys. Rev. B \textbf{92}, 014425 (2015).

\bibitem{Kus15} P. Kushwaha, V. Sunko, P. J. W. Moll, L. Bawden, J. M. Riley, N. Nandi, H. Rosner, M. P. Schmidt, F. Arnold, E. Hassinger, T. K. Kim, M. Hoesch, A. P. Mackenzie, and P. D. C. King, Science Advances \textbf{1}, 1500692 (2015).

\bibitem{Daou15} R. Daou, R. Fr\'esard, S. H\'ebert, and A. Maignan, Phys. Rev. B \textbf{91}, 041113(R) (2015).

\bibitem{Oku05} T. Okuda, N. Jufuku, S. Hidaka, and N. Terada, Phys. Rev. B \textbf{72}, 144403 (2005).

\bibitem{Noz07} T. Nozaki, K. Hayashi, and T. Kajitani, J. Chem. Eng. Japn \textbf{40}, 1205 (2007). 

\bibitem{Kur06} K. Kuriyama, M. Nohara, T. Sasagawa, K. Tabuko, F. Mizokawa, K. Kimura, and H. Takagi, {\em Proc.~25th Int.~Conf.~Thermoelectrics} (IEEE, Piscataway, 2006), p.~97.

\bibitem{Mai09b} A. Maignan, V. Eyert, C. Martin, S. Kremer, R. Fr\'esard, and D. Pelloquin, Phys. Rev. B \textbf{80}, 115103 (2009).

\bibitem{Kre12} S. Kremer and R. Fr\'esard, Ann. Phys. (Berlin) \textbf{524}, 21 (2012).

\bibitem{Mek92} 
M. Mekata, N. Yaguchi, T. Takagi, S. Mitsuda, and H. Yoshizawa, J.~Magn.~Magn.~Mater. \textbf{823}, 104 (1992). 

\bibitem{Mek93} 
M. Mekata, N. Yaguchi, T. Takagi, T. Sugino, S. Mitsuda, H. Yoshizawa, N. Hosoito, and T. Shinjo, J.~Phys.~Soc.~Japan \textbf{62}, 4474 (1993). 

\bibitem{Pet00}
O. A. Petrenko, G. Balakrishnan, M. R. Lees, D. McK.~Paul, and A. Hoser, Phys.~Rev.~B \textbf{62}, 8983 (2000).

\bibitem{Kimura2006} T. Kimura, J. C. Lashley, and A. P. Ramirez, Phys. Rev. B \textbf{73}, 220401(R) (2006).

\bibitem{Ye06}
F. Ye, Y. Ren, Q. Huang, J. A. Fernandez-Baca, P. Dai, J. W. Lynn, and T. Kimura, Phys.~Rev.~B \textbf{73},  220404(R) (2006).

\bibitem{Eye08b}
V. Eyert, R. Fr\'esard, and A. Maignan, Phys. Rev. B \textbf{78}, 052402 (2008).

\bibitem{Soda2009} M. Soda, K. Kimura, T. Kimura, M. Matsuura, and K. Hirotam, J. Phys. Soc. Jpn. \textbf{78}, 124703 (2009).

\bibitem{Mai09a} 
A. Maignan, C. Martin, R. Fr\'esard, V. Eyert, E. Guilmeau, S. H\'ebert, M. Poienar, and D. Pelloquin, Solid Stat. Comm. \textbf{149}, 962 (2009).

\bibitem{Jia12}
J. Xue-Fan, L. Xian-Feng, W. Yin-Zhong, and H. Jiu-Rong, Chin. Phys. B {\bf 21}, 077502 (2012).

\bibitem{Mazin2007}
I. I. Mazin, Phys. Rev. B \textbf{75}, 094407 (2007).

\bibitem{Haraldsen2010} J. T. Haraldsen, F. Ye, R. S. Fishman, J. A. Fernandez-Baca, Y. Yamaguchi, K. Kimura, and T. Kimura, Phys. Rev. B \textbf{82}, 020404(R) (2010).

\bibitem{Kadowaki} H. Kadowaki, H. Kikuchi, and Y. Ajiro, J. Phys.: Condens. Matter {\bf 2}, 4485 (1990).

\bibitem{Soda2010} M. Soda, K. Kimura, T. Kimura, and K. Hirota, Phys. Rev. B {\bf 81}, 100406(R) (2010).

\bibitem{Frontzek2012} M. Frontzek, G. Ehlers, A. Podlesnyak, H. Cao, M. Matsuda, O. Zaharko, N. Aliouane, S. Barilo, and S. V. Shiryaev, J. Phys.: Condens. Matter {\bf 24}, 016004 (2012).

\bibitem{Poienar2009} M. Poienar, F. Damay, C. Martin, V. Hardy, A. Maignan, and G. Andr\'e, Phys. Rev. B \textbf{79}, 014412 (2009).

\bibitem{Kimura2009} K. Kimura, T. Otani, H. Nakamura, Y. Wakabayashi, and T. Kimura, J. Phys. Soc. Jpn. \textbf{78}, 113710 (2009).

\bibitem{Seki} S. Seki, Y. Onose, and Y. Tokura, Phys. Rev. Lett. \textbf{101}, 067204 (2008).

\bibitem{Kimura2008} K. Kimura, H. Nakamura, K. Ohgushi, and T. Kimura, Phys. Rev. B \textbf{78}, 140401(R) (2008).

\bibitem{Tokura} Y. Tokura and S. Seki, Adv. Mater. {\bf 22}, 1554 (2010).

\bibitem{Tokura_Review} Y. Tokura, S. Seki, and N. Nagaosa, Rep. Prog. Phys. {\bf 77}, 076501 (2014).

\bibitem{Noriki2014} N. Terada, J. Phys.: Condens. Matter \textbf{26}, 453202 (2014).

\bibitem{Kaplan_Mahanti} T. A. Kaplan and S. D. Mahanti, Phys. Rev. B \textbf{83}, 174432 (2011).

\bibitem{rspt-book} J. M. Wills, O. Eriksson, M. Alouani, and D. L. Price, in \textit{Electronic Structure and Physical Properties of Solids, Lecture Notes in Physics}, Vol. 535, edited by H. Dreysse (Springer Berlin Heidelberg, 2000) p. 148-167.

\bibitem{crstruc} Y. Ono, K. I. Satoh, T. Nozaki, and T. Kajitani, Jpn. J. Appl. Phys. {\bf 46}, 1071 (2007).

\bibitem{lsdau} A. I. Liechtenstein, V. I. Anisimov, and J. Zaanen, Phys. Rev. B {\bf 52}, R5467(R) (1995).

\bibitem{FLL-DC} M. T. Czyzyk and G. A. Sawatzky, Phys. Rev. B {\bf 49}, 14211 (1994).

\bibitem{lichtenstein} A. I. Liechtenstein, M. I. Katsnelson, V.P. Antropov, and V. A. Gubanov, J.~Magn.~Magn.~Mater. {\bf 67}, 65 (1987).

\bibitem{exch2} M. I. Katsnelson and A.I. Lichtenstein, Phys. Rev. B {\bf 61}, 8906 (2000).

\bibitem{rspt-jijs} Y. O. Kvashnin, O. Gr\aa n\"as, I. Di Marco, M. I. Katsnelson, A. I. Lichtenstein, and O. Eriksson, Phys. Rev. B {\bf 91}, 125133 (2015).

\bibitem{MC} D.P. Landau and K. Binder, \textit{A Guide to Monte Carlo Simulations in Statistical Physics} (Cambridge University Press, Cambridge, England, 2008).

\bibitem{Metropolis} N. Metropolis, A. W. Rosenbluth, M. N. Rosenbluth, A. H. Teller, and E. Teller, J. Chem. Phys. {\bf 21}, 1087 (1953).

\bibitem{T.S.MC} U. Nowak, R. W. Chantrell, and E. C. Kennedy, Phys. Rev. Lett. \textbf{84}, 163 (2000).

\bibitem{Yamaguchi2010} H. Yamaguchi, S. Ohtomo, S. Kimura, M. Hagiwara, K. Kimura, T. Kimura, T. Okuda, and K. Kindo, Phys. Rev. B \textbf{81}, 033104 (2010).

\bibitem{Ahmed} Note that systems larger than 90$\times$90$\times$2 unit cells require more than 12.5 days of simulation which is not accessible at the super-computer of CRIANN.  

\bibitem{Okuda2013} T. Okuda, R. Kajimoto, M. Okawa, and T. Saitoh, Int. J. Mod. Phys. B \textbf{27}, 1330002 (2013).

\bibitem{frustration1} A. P. Ramirez, Annu. Rev. Mater. Sci. {\bf24}, 453 (1994).

\bibitem{frustration2} J. E. Greedan, J. Mater. Chem. \textbf{11}, 37 (2001). 

\bibitem{Kimura2009_PRL} K. Kimura, H. Nakamura, S. Kimura, M. Hagiwara, and T. Kimura, Phys. Rev. Lett. \textbf{103}, 107201 (2009).

\end{thebibliography}
\end{document}